\documentclass[12pt]{article}
\usepackage{amsmath,amssymb,amsfonts, amsbsy, epsfig, subfig}
\usepackage[usenames]{color}
\usepackage{rotating}

\newtheorem{theorem}{Theorem}
\newtheorem{corollary}{Corollary}
\newtheorem{proposition}{Proposition}
\newtheorem{lemma}{Lemma}
\newtheorem{example}{Example}
\newtheorem{remark}{Remark}
\newtheorem{definition}{Definition}
\newcommand{\beq}{\begin{equation}}
\newcommand{\eeq}{\end{equation}}
\newcommand{\beas}{\begin{eqnarray*}}
\newcommand{\eeas}{\end{eqnarray*}}
\newcommand{\bea}{\begin{eqnarray}}
\newcommand{\eea}{\end{eqnarray}}
\newcommand{\bei}{\begin{itemize}}
\newcommand{\eei}{\end{itemize}}
\newcommand{\ben}{\begin{enumerate}}
\newcommand{\een}{\end{enumerate}}
\newcommand{\bet}{\begin{theorem}}
\newcommand{\eet}{\end{theorem}}
\newcommand{\bel}{\begin{lemma}}
\newcommand{\eel}{\end{lemma}}
\newcommand{\bep}{\begin{proposition}}
\newcommand{\eep}{\end{proposition}}
\newcommand{\bed}{\begin{definition}}
\newcommand{\eed}{\end{definition}}
\newcommand{\bec}{\begin{corollary}}
\newcommand{\eec}{\end{corollary}}
\newcommand{\bex}{\begin{example}}
\newcommand{\eex}{\end{example}}
\newcommand{\qed}{\quad\hbox{\vrule width 4pt height 6pt depth 1.5pt}}
\newcommand{\RR}{I\!\! R}

\newcommand{\de}{\delta}
\newcommand{\ep}{\epsilon}

\newcommand{\goto}{\rightarrow}
\newcommand{\hf}{{1 \over 2}}

\newcommand{\argmin}{\mathop{\rm arg\min}}

\def\0{\boldsymbol{0}}
\def\u{\boldsymbol{u}}
\def\U{\boldsymbol{U}}
\def\l{\boldsymbol{l}}
\def\x{\boldsymbol{x}}
\def\be{\boldsymbol{\beta}}
\def\de{\boldsymbol{\delta}}
\def\X{\boldsymbol{X}}
\def\Y{\boldsymbol{Y}}
\def\Z{\boldsymbol{Z}}
\def\S{\boldsymbol{\Sigma}}
\def\O{\boldsymbol{\Omega}}
\def\m{\boldsymbol{\mu}}

\def\pr{\textsf{P}} 
\def\ep{\textsf{E}} 

\begin{document}

\title{ A Direct Estimation Approach to Sparse Linear Discriminant Analysis\footnote { Tony Cai is
Professor of Statistics in the Department of Statistics, The Wharton
School, University of Pennsylvania, Philadelphia, PA 19104 (Email: tcai@wharton.upenn.edu). Weidong Liu is Faculty Member, Department of Mathematics and Institute of Natural Sciences, Shanghai Jiao Tong University, China and Postdoctoral Fellow, Department of Statistics, The Wharton School, University of Pennsylvania, Philadelphia, PA 19104 (Email: liuweidong99@gmail.com). The research of Tony Cai and Weidong Liu was supported in
part by NSF FRG Grant DMS-0854973.}}
\author{Tony Cai and Weidong Liu}

\date{}

\maketitle
\vspace{-1cm}

\begin{abstract}
This paper considers sparse linear discriminant analysis of high-dimensional data. In contrast to the existing methods which are based on separate estimation of the precision matrix $\O$ and the difference $\de$ of the mean vectors, we introduce a simple and effective classifier by estimating the product $\O\de$ directly through constrained $\ell_1$ minimization. The estimator can be implemented efficiently using linear programming and the resulting classifier is called the linear programming  discriminant (LPD) rule.

The LPD rule is shown to have desirable theoretical and numerical properties. It exploits the approximate sparsity of $\O\de$ and as a consequence allows cases where it can still perform well even when $\O$ and/or $\de$ cannot be estimated consistently.  Asymptotic properties of the LPD rule are investigated and consistency and rate of convergence results are given. The LPD classifier has superior finite sample performance and significant computational advantages over the existing methods that require separate estimation of $\O$ and $\de$. The LPD rule is also applied to analyze real datasets from lung cancer and leukemia studies. The classifier performs favorably in comparison to existing methods.
\end{abstract}


\newpage

\section{Introduction}
Classification is an important problem which has been well studied  in the classical  low-dimensional setting. In particular, linear discriminant analysis (LDA), which uses a linear combination of features as the criterion for classification,  has been shown to perform well and enjoy certain optimality as the sample size tends to infinity while the dimension is fixed.
Consider two $p$-dimensional normal distributions  $N(\m_{1},\S)$ (class 1) and $N(\m_{2},\S)$ (class 2) with the same covariance matrix. Let $\Z$ be a random vector that is drawn from one of these two distributions with equal prior probabilities.  The goal of classification is to determine from which class $\Z$ is drawn. The problem is simple in the ideal setting where the parameters $\m_{1}$, $\m_{2}$, and $\S$  are known in advance. In this case,
Fisher's  linear discriminant rule
\begin{eqnarray}\label{f}
\psi_{\boldsymbol{F}}(\boldsymbol{Z})=I\{(\boldsymbol{Z}-\boldsymbol{\mu})^{'}\boldsymbol{\O}\de \geq 0\},
\end{eqnarray}
where $\boldsymbol{\mu}=(\boldsymbol{\mu}_{1}+\boldsymbol{\mu}_{2})/2$, $\de=\m_{1}-\m_{2}$ and $\O=\S^{-1}$,  classifies $\Z$ into class 1 if and only if $\psi_{\boldsymbol{F}}(\boldsymbol{Z})=1$. This classifier is the Bayes rule with equal prior probabilities for the two classes and is thus optimal in such an ideal setting.

Fisher's rule can be used to serve as an oracle benchmark, but it is typically not directly applicable in real data analysis as the parameters are usually unknown and need to be estimated from the samples. It is a standard practice to separately estimate
$\O$ and $\de$ and then plug the estimates into (\ref{f}) to construct a classifier. Let
$\{\X_{k}; 1\leq k\leq n_{1}\}$ and $\{\Y_{k}; 1\leq k\leq n_{2}\}$ be independent and identically distributed random samples from $N(\m_{1},\S)$ and $N(\m_{2},\S)$ respectively. The classical estimates of $\m_{1},\m_{2}$  and $\O$ in the low-dimensional setting  are the sample means $\bar \X$ and $\bar \Y$ and the inverse sample covariance matrix $\hat{\S}_n^{-1}$.
Plugging these estimates into (\ref{f}) results in
$\hat{\psi}_{\boldsymbol{F}}(\boldsymbol{Z})$, the empirical version of $\psi_{\boldsymbol{F}}(\boldsymbol{Z})$. Theoretical properties of $\hat{\psi}_{\boldsymbol{F}}(\boldsymbol{Z})$ has been well studied when $p$ is fixed and can be found, for example,  in Anderson (2003).

With dramatic advances in technology, high-dimensional data are now routinely collected in a wide range of applications and classification for these data has drawn considerable recent attention. Examples include  genomics, functional magnetic resonance imaging, risk management and web search problems.
In the high-dimensional settings, the standard LDA performs poorly and can even fail completely. For example, Bickel and Levina (2004) showed that the LDA can be no better than random guessing when $p/(n_{1}+n_{2})\goto \infty$. In such a setting, the sample covariance matrix $\hat{\S}_n$ is singular and its inverse is not well defined.  One natural  remedy is to use instead the generalized inverse of the sample covariance matrix.  However, such an estimate is highly biased and unstable and will lead to a classifier with poor performance when $p$ is large.
A naive method in this case is to simply ignore the dependence among the variables and replace $\S$ with the diagonal of the sample covariance matrix. This leads to the so-called naive Bayes rule, also called the independence rule; see Bickel and Levina (2004). Assuming that the difference $\de$  is sparse, Fan and Fan (2008)  proposed the features annealed independence rule which applies the naive independence rule to a set of selected important features of $\de$ that are chosen by thresholding. This rule ignores the correlations between the variables and can be inefficient. See Section \ref{discussion.sec} for further discussions.

In the high-dimensional setting, regularity conditions on $\O$ (or $\S$) and $\de$ are needed to ensure that they can be estimated consistently. The most commonly used structural assumptions are that  $\O$ (or $\S$) and $\de$ are sparse. Under such assumptions,  $\O$ and $\de$ are estimated separately and are then plugged into the Fisher's rule (\ref{f}).  Assuming the covariance matrix $\S$  and the difference $\de$ are sparse,  Shao, et al. (2011) used the thresholding procedures for estimating $\S$ and $\de$.  More commonly in applications the sparsity assumption is on the precision matrix $\O$ instead of $\S$. In such a setting, Rothman, et al. (2008) used the Glasso estimator for $\O$ in (\ref{f}).  Witten and Tibshirani (2009) proposed the scout procedure for classification in which they replaced $\O$ with a shrunken estimate.
See also  Friedman (1989), Tibshirani, et al. (2002), Guo, et al. (2007),  Wu, et al. (2009), and Hall, et al. (2009) and the reference therein.

A simple but important observation is that the Fisher's  discriminant rule (\ref{f}) depends on $\O$ and $\de$ only through their product $\O\de$. In the present paper, we shall show that the product $\O\de$ can be estimated  directly and efficiently, even when $\O$ and/or $\de$ cannot be well estimated individually. We introduce the following direct estimation method for sparse linear discriminant analysis by estimating $\O\de$ through a constrained $\ell_1$ minimization method.
Specifically, we propose to estimate $\be^{*}:=\O\de$ by
\[
\hat \be \in \argmin_{\be\in \RR^p} \{|\be|_{1}\quad\mbox{subject to}\quad |\hat{\S}_n\be-(\bar{\X}-\bar{\Y})|_{\infty}\leq \lambda_{n}\},
\]
where $\lambda_{n}$ is a tuning parameter, and classify $\Z$ to class $1$ if and only if
\[
(\Z-\hat{\m})^{'}\hat{\be}\geq 0,
\]
where $\hat{\m}=(\bar{\X}+\bar{\Y})/2.$ The estimator $\hat \be$ can be implemented easily using linear programming. The resulting classification procedure is thus called the {\it linear programming  discriminant (LPD) rule}.
The LPD  rule is data-driven and easy to implement. It has significant computational advantage over the existing methods that require separate estimation of $\O$ (or $\S$) and $\de$, because it only requires the estimation of a $p$-dimensional vector via linear programming instead of the estimation of the inverse of a $p\times p$ covariance matrix.

Both the theoretical and numerical properties of the LPD rule are studied in this paper. The LPD rule performs well when $\O\de$ is approximately sparse, which is a weaker and more flexible assumption than that both $\O$ and $\de$ are sparse. In particular, under this assumption the precision matrix $\O$ is not required to be sparse and may not be consistently estimable. The asymptotic properties of the LPD rule are investigated and consistency and rate of convergence results are given.  In addition to the Gaussian case, extensions to the non-Gaussian distributions are also considered.
Numerical performance of the LPD classifier is investigated using both simulated and real data. A simulation study is carried out and the numerical results show that the LPD rule has superior finite sample performance in comparison to several other classifiers. It significantly outperforms the alternative methods in terms of the average misclassification rate. The LPD rule is also applied to the analysis of  real datasets from lung cancer and leukemia studies and performs favorably in comparison to existing
methods.

The rest of the paper is organized as follows. Section \ref{method.sec} introduces a constrained $\ell_1$ minimization method for the direct estimation of  $\O\de$ which leads to the LPD  classification rule. Section \ref{theory.sec} investigates the asymptotic properties of the LPD rule in the Gaussian setting. Extensions to non-Gaussian distributions are given in Section \ref{extension.sec}. Section \ref{numerical.sec} first discusses the linear programming implementation of the LPD classifier, and then investigates the numerical performance of the LPD rule by simulations and by applications to lung cancer and leukemia datasets. Discussions of our results and other related work are given in Section \ref{discussion.sec}. The main results are proved  in Section \ref{proof.sec}.

\section{Classification via direct estimation of $\O\de$}
\label{method.sec}

In this section we introduce a constrained $\ell_1$ minimization method for estimating the product  $\O\de$ directly.  It will be shown in Sections \ref{theory.sec} - \ref{numerical.sec} that the resulting classification rule enjoys desirable properties  theoretically, computationally, and numerically. For ease of presentation, we shall focus on the Gaussian case in this section and Section \ref{theory.sec}. The non-Gaussian case is considered in Section \ref{extension.sec}. We begin by reviewing basic notation and definitions.

For a vector $\be=(\beta_{1},\ldots,\beta_{p})^{'}\in \RR^p$, define the $\ell_{0}$ norm by $|\be|_{0}=\sum_{j=1}^{p}I\{\beta_{j}\neq 0\}$; the $\ell_{q}$ norm by  $|\be|_{q}=(\sum_{i=1}^{p}|\beta_{i}|^{q})^{1/q}$ for $1\le q \le \infty$ with the usual modification for $q=\infty$. The vector $\beta$ is called $k$-sparse if it has at most $k$ nonzero entries. For a matrix $\O=(\omega_{ij})_{p\times p}$,  the matrix $1$-norm is defined to be the maximum absolute column sum, $\|\Omega\|_{L_{1}}=\max_{1\leq j\leq p}\sum_{i=1}^{p}|\omega_{ij}|$. For a  matrix $\O$, we say $\O$ is $k$-sparse if each row/column has at most $k$ nonzero entries.
For two  sequences of real numbers $\{a_{n}\}$ and $\{b_{n}\}$, write $a_{n} = O(b_{n})$ for $n\geq 1$ if there exists a constant $C$ such that $|a_{n}| \leq C|b_{n}|$, write $a_{n} = o(b_{n})$ if $\lim_{n\rightarrow\infty}a_{n}/b_{n} = 0$, and write $a_{n}\asymp b_{n}$ if there are positive constants $c$ and $C$ such that $c\leq a_{n}/b_{n}\leq C$ for all $n\ge 1$.

Recall that $\{\X_{k}; 1\leq k\leq n_{1}\}$ and $\{\Y_{k}; 1\leq k\leq n_{2}\}$ are independent and identically distributed random samples from $N(\m_{1},\S)$ and $N(\m_{2},\S)$ respectively. Set
 \beq\label{a0}
\bar{\X}=\frac{1}{n_{1}}\sum_{i=1}^{n_{1}}\X_{i},\quad \bar{\Y}=\frac{1}{n_{2}}\sum_{i=1}^{n_{2}}\Y_{i},
\quad \hat{\de}=\bar{\X}-\bar{\Y},\quad\hat{\m}=(\bar{\X}+\bar{\Y})/2.
\eeq
Denote the sample covariance matrices by
\[
\hat{\S}_{\X}=\frac{1}{n_{1}}\sum_{i=1}^{n_{1}}(\X_{i}-\bar{\X})(\X_{i}-\bar{\X})^{'},\quad \hat{\S}_{\Y}=\frac{1}{n_{2}}\sum_{i=1}^{n_{2}}(\Y_{i}-\bar{\Y})(\Y_{i}-\bar{\Y})^{'},
\]
and set
\[
\hat{\S}_{n}=\frac{1}{n}(n_{1}\hat{\S}_{\X}+n_{2}\hat{\S}_{\Y}),
\]
 where $n=n_{1}+n_{2}$.

As mentioned in the introduction, most of the classification methods in the literature involve separate estimation of  the unknown precision matrix $\O=\S^{-1}$ and the difference of the means $\de$ in the Fisher's rule (\ref{f}). In the high-dimensional setting, the sample covariance matrix $\hat{\S}_n$ is typically not invertible and regularity conditions are needed in order to be possible to construct good estimators. It should be noted that accurate estimation of a large sparse precision matrix is a difficult and computationally costly problem itself.  See, e.g., Ravikumar, et al. (2008), Yuan (2009), and Cai, Liu and Luo (2011).

It is clear that the Fisher's rule (\ref{f}) depends on $\O$ and $\de$ only through their product $\O\de$. We now introduce a constrained $\ell_1$ minimization method to directly  estimate the product $\O\de$ by exploiting the (approximate) sparsity of $\O\de$. We should note here that the sparsity of $\O\de$ is a weaker and more flexible condition than the  sparsity of both $\O$ and $\de$.  In particular, it does not require the precision matrix $\O$ to be sparse. See Remark \ref{sparsity.remark} below for more discussions. Specifically, we propose to estimate $\be^{*}:=\O\de$ by the solution to the following optimization problem:
\begin{eqnarray}\label{a1}
\hat \be \in \argmin_{\be\in \RR^p} \{|\be|_{1}\quad\mbox{subject to}\quad |\hat{\S}_n\be-(\bar{\X}-\bar{\Y})|_{\infty}\leq \lambda_{n}\},
\end{eqnarray}
where $\lambda_{n}$ is a tuning parameter which will be specified later. The constrained $\ell_1$ minimization method (\ref{a1}) is known to be an effective way for reconstructing sparse signals. The readers are referred to Donoho, et al.\ (2006) and Cand\`es and Tao (2007) for more details on the $\ell_1$ minimization methods for sparse signal recovery. We shall show that the direct estimate leads to a classifier that is more effective and efficient than those based on estimating  $\O$ and $\de$ separately.

Given  the solution $\hat{\be}$ to (\ref{a1}), we propose the following classification rule: classify $\Z$ to class $1$ if and only if
\begin{eqnarray}\label{a2}
(\Z-\hat{\m})^{'}\hat{\be}\geq 0.
\end{eqnarray}
The optimization problem (\ref{a1}) can be cast as a linear program. We shall call the discriminant in (\ref{a2}) the  {\it Linear Programming Discriminant (LPD)} and the classification rule (\ref{a2})  the LPD rule.

The motivation behind the constrained $\ell_1$ minimization method (\ref{a1}) for estimating $\be^{*}=\O\de$ directly can be easily seen as follows. Note that $\be^{*}$ is the solution to the equation $\S\be-\de =\0$. When $\S$ and $\de$ are unknown, they are replaced by their respective sample versions $\hat \S_n$ and $\hat \de = \bar{\X}-\bar{\Y}$. We then seek the most sparse solution within the feasible set
\[
\{\be: \; |\hat{\S}_n\be-(\bar{\X}-\bar{\Y})|_{\infty}\leq \lambda_{n}\}
\]
to account for the variability in  $\hat \S_n$ and $\hat \de$.
The convex relaxation of using $\ell_1$ minimization in place of $\ell_0$ minimization is a standard technique in sparse signal recovery.  We shall show in the next sections that the resulting classification rule (\ref{a2}) has desirable properties both asymptotically and numerically. The $\ell_1$ minimization method (\ref{a1}) works well when $\O\de$ is approximately sparse. It thus allows the case where $\O$ itself is not sparse. In other words, it is possible to classify $\Z$ with accuracy using the classifier (\ref{a2}) even when $\O$ cannot be estimated consistently.

 In addition to its good performance in terms of classification accuracy, the classifier given in (\ref{a2}) also enjoys significant computational advantages over existing methods that require separate estimation of $\O$ and $\de$. This can be seen at an intuitive level. There is only $p$ parameters in $\O\de$, while one needs to estimate $p^{2}/2$ parameters if $\O$ and $\de$ are estimated separately.  More discussions on the computational issues will be given in Section \ref{numerical.sec}.

\begin{remark} \label{sparsity.remark}
{\rm
It is easy to see that if $\de$ is $k_1$-sparse and $\O$ is $k_2$-sparse, then $\O\de$ is at most $k_1k_2$-sparse.
Furthermore, the sparsity of $\O\de$  does not require $\O$ being sparse. Suppose $\de$ is $k_1$-sparse and without loss of generality assume the nonzeros are among the first $k_1$ coordinates. (In general we can always re-order the rows/columns of $\O$ accordingly.)  So, $\de$ can be written as
\begin{eqnarray*}
\de=\left(
\begin{array}{cc}
 \de_{1} \\ \0
\end{array}
\right)
\end{eqnarray*}
where $\de_{1}$ is a $k_1$-dimensional vector.  Write $\O$ as
 \begin{eqnarray*}
\O =\left(
\begin{array}{cc}
 \O_{11} & \O'_{21} \\ \O_{21} & \O_{22}
\end{array}
\right),
\end{eqnarray*}
where  $ \O_{11}$ is $k_1\times k_1$, $ \O_{21}$ is $(p-k_1)\times k_1$, and $ \O_{22}$ is $(p-k_1)\times (p-k_1)$.  Then the sparsity of $\O\de$ does not depend on the submatrix $ \O_{22}$ at all. $\O\de={\O_{11}\de_1\choose \O_{21}\de_1}$ is sparse if $ \O_{21}$ is sparse. In particular, if there are at most $k_{2}$ nonzero elements on each column of $\O_{21}$, then $\O\de$ is $k_{1}(k_{2}+1)$ sparse.
No condition on $\O_{22}$ is needed. In general, it is not possible to consistently estimate $\O$ under the spectral norm without regularity conditions on $\O_{22}$.
The consistency of $\hat \O$ was required by Shao, et al. (2011) through the invertibility of the estimated covariance matrix $\hat \S$ and for the good asymptotic performance of the resulting classification rule. In fact, even when $\O$ is the identity matrix, joint estimation of $\O\de$ by (\ref{a1}) may lead to a better misclassification rate than estimating $\O$ and $\de$ separately as in Shao, et al. (2011). See Remark 5 for more details.

Finally we note that there are also cases that  neither $\O$ nor $\de$ is sparse, but $\O\de$ is. For example, if $\de=(\sigma_{11},\ldots,\sigma_{p1})^{'}$, the first column of $\S$, then $\O\de=(1,0,\ldots,0)^{'}$. Hence the sparsity on $\O\de$ is more flexible than assuming both $\O$ and $\de$ are  sparse.}
\end{remark}

\section{Asymptotic properties}
\label{theory.sec}

We now turn to the theoretical properties of the LPD classifier given in (\ref{a2}). Both consistency and convergence rate results are given. We shall focus on the Gaussian case in this section. Extensions to the non-Gaussian case are discussed in Section \ref{extension.sec} and numerical performance of the classifier will be considered in Section \ref{numerical.sec}.

The  misclassification rate of the Fisher's rule (see, e.g., Anderson (2003)) is
\begin{eqnarray}\label{a3-3}
R:=1-\Phi(\Delta^{1/2}_{p}) \quad \mbox{with}\quad \Delta_{p}=\de^{'}\O\de,
\end{eqnarray}
which is the best possible performance in the ideal setting where all the parameters $\m_{1}$, $\m_{2}$, and $\S$ are known in advance. This can serve as an oracle benchmark for the performance of any data-driven classifier based on the samples $\{\X_k\}$ and $\{\Y_k\}$.

It is not difficult to calculate  that, given the samples $\{\X_{k}\}$ and $\{\Y_{k}\}$, the  conditional misclassification rate of the LPD rule  is
\[
R_{n}:=1-\hf \Phi\Big{(}-\frac{(\hat{\m}-\m_{1})^{'}\hat{\be}}{(\hat{\be}^{'}\S\hat{\be})^{1/2}}\Big{)}- \hf \Phi\Big{(}\frac{(\hat{\m}-\m_{2})^{'}\hat{\be}}{(\hat{\be}^{'}\S\hat{\be})^{1/2}}\Big{)}
\]
where $\hat \be$ is given in (\ref{a1}).
The performance of the LPD rule can then be naturally measured by the difference (or ratio) between $R_n$ and the Bayes misclassification rate $R$.
In this section we will study the difference and ratio between $R_{n}$ and $R$. To this end,
we need to introduce some conditions.\\\vspace{-5mm}

\noindent{\bf (C1).}  $n_1\asymp n_2$, $\log p\leq n$, $c^{-1}_{0}\leq\lambda_{\min}(\S)\leq \lambda_{\max}(\S)\leq c_{0}$ for some constant $c_{0}>0$ and $\Delta_{p}\geq c_{1}$ for some $c_{1}>0$.\\\vspace{-5mm}

Here we assume that the two samples are of comparable sizes and the eigenvalues of the covariance matrix $\S$ are bounded from below and above. These are commonly used conditions in the high dimensional setting. In addition, we also assume $\Delta_{p}$ is bounded away from zero. If $\Delta_{p}\rightarrow 0$, then it can be seen easily from (\ref{a3-3}) that even the oracle rule is no better than random guessing.

Our first result is on the consistency of $R_{n}$.

\begin{theorem}\label{th1-0} Let $\lambda_{n}=C\sqrt{\Delta_p \log p/n}$ with $C>0$ being a sufficiently large constant. Suppose (C1) holds and
\begin{eqnarray}\label{c1-0}
|\O\de|_{0}=o\left(\sqrt{\frac{n}{\log p}}\right).
\end{eqnarray}
Then we have as $n\rightarrow\infty$ and $p\rightarrow\infty$,
\begin{eqnarray}\label{th1-1-0}
R_{n}-R\rightarrow 0
\end{eqnarray}
in probability.
\end{theorem}
This theorem shows that the LPD rule is consistent when $\O\de$ is sparse. In practice, the value of the tuning parameter $\lambda_n$ is chosen by cross-validation. See Section \ref{numerical.sec} for further discussions on the implementation of the LPD rule.

\begin{remark}{\rm
As mentioned earlier, the condition $|\O\de|_{0}=o\Big{(}\sqrt{n/\log p}\Big{)}$ does not require $\O$ to be sparse. Therefore, by estimating $\O\de$ directly,  we do not need a consistent estimate for $\O$ or $\S$ under the spectral norm to get the asymptotically optimal misclassification rate. In contrast, consistent estimation of $\O$ is required by Shao, et al. (2011). A basic condition in Shao, et al. (2011) is that $\S=:(\sigma_{ij})_{p\times p}$ is (approximately) sparse with the sparsity $s_{0}(p)$ of $\S$ satisfying $s_{0}(p)(\log p/n)^{(1-q)/2}=o(1)$ and $\max_{1\leq i\leq p}\sum_{j=1}^{p}|\sigma_{ij}|^{q}\leq s_{0}(p)$ for $0\leq q<1$.  It follows from Cai and Zhou (2010) on the minimax rate of convergence for estimating sparse covariance matrices, this condition is necessary  for the consistency under the spectral norm.
 }
 \end{remark}

 Theorem \ref{th1-0} can be extended to a more general setting where $\O\de$ is only approximately sparse. To state this result, we first relax the condition (C1) as follows.\\\vspace{-5mm}

\noindent{\bf (C2).} $n_1\asymp n_2$,  $\log p\leq n$,  $\max_{1\leq i\leq p}\sigma_{ii}\leq K$ for some constant $K>0$ and $\Delta_{p}\geq c_{1}$ for some constant $c_{1}>0$.

\begin{theorem}\label{th1} Let $\lambda_{n}=C\sqrt{ \Delta_p \log p/n}$ with $C$ being a sufficiently large constant. Suppose (C2) holds and
\beq\label{c1}
\frac{|\O\de|_{1}}{\Delta^{1/2}_{p}}+\frac{|\O\de|^{2}_{1}}{\Delta^{2}_{p}}=o\left(\sqrt{\frac{n}{\log p}}\right).
\eeq
Then we have as $n\rightarrow\infty$ and $p\rightarrow\infty$,
\begin{eqnarray}\label{th1-1}
R_{n}-R\rightarrow 0
\end{eqnarray}
in probability.
\end{theorem}

\begin{remark}{\rm
It follows from the Cauchy-Schwarz inequality and (C1),
\[
|\O\de|^{2}_{1}\leq |\O\de|_{0}|\O\de|^{2}_{2}\leq c^{2}_{0}|\O\de|_{0}|\de|^{2}_{2}
\]
and $\Delta_{p}\geq c^{-1}_{0}|\de|^{2}_{2}$. Thus (\ref{c1-0}) implies (\ref{c1}). The condition (\ref{c1-0}) can be further relaxed if the minimum magnitude of the nonzero elements of $\O\de$ is relatively large.
Let $S=\{i: (\O\de)_{i}\neq 0\}$. If $\min_{i\in S}|(\O\de)_{i}|\geq C(\log p/n)^{1/4}$, then
a sufficient condition of (\ref{c1}) is $|\O\de|_{0}=o(n/\log p)$. Condition (\ref{c1}) allows the case where $\O\de$ is only approximately sparse with many small entries.
}
\end{remark}

\begin{remark}{\rm
The condition $\max_{1\leq i\leq p}\sigma_{ii}\leq K$ can be relaxed. Let $K_p:=\max_{1\leq i\leq p}\sigma_{ii}$ and $\lambda_{n}=C\sqrt{K_{p} \Delta_p \log p/n}$. Theorem \ref{th1} still holds under the condition
\[
\frac{|\O\de|_{1}}{\Delta^{1/2}_{p}}+\frac{|\O\de|^{2}_{1}}{\Delta^{2}_{p}}=o\left(\sqrt{\frac{n}{K_p \log p}}\right).
\]
Here $K_{p}$ can grow and may tend to infinity as $p\goto \infty$.
}
\end{remark}


Theorems \ref{th1-0} and \ref{th1} provide the consistency results for the LPD rule.  Consistency is important, but the fact $R_{n}-R\rightarrow 0$ does not give a detailed description of the properties of a classifier. For example, when the Bayes misclassification rate $R\rightarrow 0$, any classifier with $R_{n}\rightarrow 0$ is consistent.
Stronger results on the rate of convergence can be obtained.

\begin{theorem}\label{th2}
Let $\lambda_{n}=C\sqrt{\Delta_p \log p/n}$ with $C$ being a sufficiently large constant. Suppose (C2) holds and
\[
|\O\de|_{1}\Delta^{1/2}_{p}+|\O\de|_{1}^{2} =o\left(\sqrt{\frac{n}{\log p}}\right).
\]
Then
\begin{eqnarray*}
\frac{R_{n}}{R}-1=O\left((|\O\de|_{1}\Delta^{1/2}_{p}+|\O\de|_{1}^{2})\sqrt{\frac{\log p}{n}}\right)
\end{eqnarray*}
with probability greater than $1-O(p^{-1})$. In particular, if (C1) holds and $$|\O\de|_{0}\Delta_{p}=o\Big{(}\sqrt{\frac{n}{\log p}}\Big{)},$$ then
\begin{eqnarray*}
\frac{R_{n}}{R}-1=O\left(|\O\de|_{0}\Delta_{p}\sqrt{\frac{\log p}{n}}\right)
\end{eqnarray*}
with probability greater than $1-O(p^{-1})$.
\end{theorem}

 Theorem \ref{th2} shows that a larger $\Delta_{p}$ implies a worse convergence rate for the relative classification error $R_n/R$. This is in fact to be expected.  When $\Delta_{p}$ is large, the classification problem is easy and the Bayes misclassification rate $R$ can be very small. It then  becomes harder for any data-driven classification rule to mimic the performance of the oracle rule.

\begin{remark}{\rm
Due to the differences in setting, it is not directly comparable between our results and the results in Shao, et al. (2011). To make them comparable,  it is necessary to
assume both $\S$ and $\O$ are sparse. For simplicity, we consider the case $\S=I_{p\times p}$. Suppose that $|\de|_{1}\leq K$ for some constant $K$ and $\log p=o(n)$. Theorem  \ref{th2} shows that $R_{n}/R-1=O_{\pr}(\sqrt{\log p/n})$. Let $R_{n}^{*}$ be the conditional misclassification rate of the  SLDA rule proposed in Shao, et al. (2011). Their results show that $R^{*}_{n}/R-1=O_{\pr}(b_{n})$ with $b_{n}=(\log p/n)^{\alpha(1-g)}|\de|_{2g}^{g}$ for some $0<\alpha<1/2$ and $0\leq g<1$. It is easy to see that $b_{n}/\sqrt{\log p/n}\rightarrow\infty$. So the LPD rule outperforms the SLDA rule in this case.

}
\end{remark}

 The convergence rate in Theorem \ref{th2} can be further improved under stronger conditions.



\begin{theorem}\label{th3}
Let $\lambda_{n}=C\sqrt{\Delta_p \log p/n}$ with $C$ being a sufficiently large constant. Suppose (C2) holds and
\begin{eqnarray}\label{th3-3}
\|\O\|_{L_{1}}|\O\de|_{0}+|\O\de|_{1}\Delta^{1/2}_{p}=o\left(\sqrt{\frac{n}{\log p}}\right).
\end{eqnarray}
Then
\begin{eqnarray*}
\frac{R_{n}}{R}-1=O\left(|\O\de|_{1}\Delta^{1/2}_{p}\sqrt{\frac{\log p}{n}}\right)
\end{eqnarray*}
with probability greater than $1-O(p^{-1})$. In particular, if (C1) holds and $$\|\O\|_{L_{1}}|\O\de|_{0}+|\O\de|^{1/2}_{0}\Delta_{p}=o\Big{(}\sqrt{\frac{n}{\log p}}\Big{)},$$ then
\begin{eqnarray*}
\frac{R_{n}}{R}-1=O\left(|\O\de|^{1/2}_{0}\Delta_{p}\sqrt{\frac{\log p}{n}}\right)
\end{eqnarray*}
with probability greater than $1-O(p^{-1})$.
\end{theorem}

\section{Extensions}
\label{extension.sec}

Section \ref{theory.sec} establishes the theoretical properties of the LPD classifier in the Gaussian setting. The results can be extended to a class of non-Gaussian distributions satisfying certain moment conditions.

Let  $\X$ and $\Y$ be two $p$-dimensional random vectors satisfying
\begin{eqnarray*}
\X=\m_{1}+\U_{1}\quad\mbox{and\quad}\Y=\m_{2}+\U_{2},
\end{eqnarray*}
where $\U_{1}$ and $\U_{2}$ are independent and identically distributed random vectors with mean zero and covariance matrix $\S=(\sigma_{ij})_{p\times p}$.
Fang and Anderson (1990) showed that the Fisher's rule is still optimal when $\U_{1}$ has an elliptical distribution with zero mean and density
\begin{eqnarray}\label{c2}
c_{p}|\S|^{-1/2}f\Big{(}\u^{'}\S^{-1}\u\Big{)},
\end{eqnarray}
where $f$ is a monotone function on $[0,\infty)$ and $c_{p}$ is a normalizing constant. The optimal misclassification rate in this case  is
\begin{eqnarray*}
R=\frac{1}{2}\pr\Big{(}\U^{'}_{1}\O\de< -\de^{'}\O\de\Big{)} + \frac{1}{2}\pr\Big{(}\U^{'}_{2}\O\de\geq \de^{'}\O\de\Big{)}.
\end{eqnarray*}
As in Shao, et al. (2011), we relax the normality of $\U_{1}$ to that, for any $p$ dimensional non-random vector $\l$ with $|\l|_{2}=1$ and any $t\in R$,
\begin{eqnarray*}
\pr\Big{(}\l^{'}\O^{1/2}\U_{1}\leq t\Big{)}=:\Psi(t)
\end{eqnarray*}
is a continuous distribution function symmetric about $0$ and does not depend on $\l$. The elliptical distributions (such as (\ref{c2})) and the multivariate scale mixture of normals satisfy this condition. The  conditional classification error of the LPD rule (\ref{a2}) given $\{\X_{k}\}$ and
$\{\Y_{k}\}$ is
\[
R_{n}:=1-\frac{1}{2} \Psi\Big{(}-\frac{(\hat{\m}-\m_{1})^{'}\hat{\be}}{(\hat{\be}^{'}\S\hat{\be})^{1/2}}\Big{)}
-\frac{1}{2} \Psi\Big{(}\frac{(\hat{\m}-\m_{2})^{'}\hat{\be}}{(\hat{\be}^{'}\S\hat{\be})^{1/2}}\Big{)}.
\]
To obtain the
convergence rate for $R_{n}$,  we shall impose an additional condition: for any $x<0$ and $|\delta|\leq 1$,
\begin{eqnarray}\label{a12}
\Big{|}\frac{\Psi(x+\delta)}{\Psi(x)}-1\Big{|}\leq c_{1}|\delta|(|x|+1)e^{c_{2}|x\delta|}
\end{eqnarray}
for some positive constants $c_{1}, c_{2}$ which do not depend on $x$ and $\delta$. Note that the distribution with density function $p(x)=c_{3}(1+|x|)^{-w}e^{-c_{4}|x|^{\varphi}}$ satisfies (\ref{a12}),
where  $c_{3}$ and $c_{4}$ are positive constants, $\varphi$ and $w$ are constants with $0< \varphi\leq 2$, $w\geq 0$, or $\varphi=0$, $w>1$.

The moment conditions are divided into two cases: the  sub-Gaussian-type tails and the polynomial-type tails. Let $\U_{1}=:(U_{1},\ldots,U_{p})^{'}$ and $U_{\de}=\U^{'}_{1}\O\de/\Delta^{1/2}_{p}$.
Note that $U_{\de}$ is a standardized random variable with zero mean and unit variance.

\noindent{\bf (C3). (Sub-Gaussian-type tails)}  Suppose that $\log p\leq n$ and there exist some constants
 $\eta>0$  and $K_{1}>0$ such that
\begin{eqnarray}\label{c3-4}
\ep \exp\Big{(}\eta U^{2}_{\de}\Big{)}\leq K_{1},\quad\mbox{and\quad} \ep \exp\Big{(}\eta U^{2}_{i}/\sigma_{ii}\Big{)}\leq K_{1}~~~\mbox{for all
$i$.}
\end{eqnarray}

\noindent{\bf (C4). (Polynomial-type tails)} Suppose that for some $\gamma,
c_{1}>0$,  $ p\leq c_{1}n^{\gamma}$, and for some $\epsilon>0$
\begin{eqnarray}\label{c3-3}
\ep|U_{\de}|^{4\gamma+4+\epsilon}\leq K_{1} \quad\mbox{and\quad}\ep|U_{i}/\sigma_{ii}^{1/2}|^{4\gamma+4+\epsilon}\leq K_{1}~~~\mbox{for all $i$.}
\end{eqnarray}

\begin{theorem}\label{th4} (i). Assume that the conditions in Theorem \ref{th1} hold. By replacing the normality with  elliptical distributions satisfying (C3) or (C4),
we have as $n\rightarrow\infty$ and $p\rightarrow\infty$,
\begin{eqnarray}\label{th1-1}
R_{n}-R\rightarrow 0 \quad \mbox{\rm in probability.}
\end{eqnarray}
(ii). Under the conditions in Theorem \ref{th2}, (\ref{a12}) and (C3) (or (C4)),
\begin{eqnarray*}
\frac{R_{n}}{R}-1=O\left((|\O\de|_{1}\Delta^{1/2}_{p}+|\O\de|_{1}^{2})\sqrt{\frac{\log p}{n}}\right)
\end{eqnarray*}
with probability greater than $1-O(p^{-1}+n^{-\epsilon/8})$.

\end{theorem}

Similarly, Theorem \ref{th3} remains valid if  the normality assumption is replaced by  elliptical distributions satisfying (C3) or (C4). For reasons of space, we do not restate the results here.

\section{Numerical Investigation}
\label{numerical.sec}

We now turn to the numerical performance of the LPD rule using both simulated and real data.   We begin in Section \ref{implementation.sec} with a discussion on the implementation of the classifier using linear programming and the selection of the tuning parameter $\lambda_n$ through cross-validation. Section \ref{simulation.sec} presents simulation results and comparisons with other methods including oracle features annealed independence rule (OFAIR) (the support of $\de$ is assumed to be known),  the nearest shrunken centroids method (NSC) proposed by Tibshirani, et al. (2002), the sparse linear discriminant (SLD) introduced in Shao, et al. (2011),  the Naive-Bayes rule (Naive-LDA), the LDA rule with a generalized inverse (GLDA)  as well as the oracle Fisher's rule (Oracle).  The applications of the LPD rule to the analysis of a lung cancer dataset and a leukemia dataset are given in Section \ref{real.data.sec}.

\subsection{Implementation of LPD}
\label{implementation.sec}

Recall that the estimate of $\be^{*}=\O\de$ is obtained by solving the constrained $\ell_1$ minimization problem
\[
\hat \be \in \argmin_{\be\in \RR^p} \{|\be|_{1}\quad\mbox{subject to}\quad |\hat{\S}_n\be-(\bar{\X}-\bar{\Y})|_{\infty}\leq \lambda_{n}\}.
\]
This optimization problem is convex, and  can easily be recast as the following linear program,
\begin{equation}
  \begin{split}\label{eq:lin}
    &\min  \sum_{j=1}^p u_j \\
    \text{subject to: } & - \beta_j \le u_j \text{ for all }1\le j \le
    p\\
    & + \beta_j \le u_j \text{ for all }1 \le j \le
    p\\
    & -\hat{\boldsymbol{\sigma}}_k^{'} \boldsymbol{\beta} + \hat{\delta}_{k} \le \lambda_n \text{ for all }1\le
    k \le
    p \\
    & + \hat{\boldsymbol{\sigma}}_k^{'} \boldsymbol{\beta} -\hat{\delta}_{k}  \le \lambda_n \text{ for all
    }1\le k \le p,
  \end{split}
\end{equation}
where $(\hat{\delta}_{1},\ldots,\hat{\delta}_{p})^{'}:=\hat{\de}$ and $(\hat{\boldsymbol{\sigma}}_{1},\ldots,\hat{\boldsymbol{\sigma}}_{p}):=\hat{\S}_{n}$.

This linear programming implementation is similar to that of the Dantzig selector in high-dimensional linear regression.  See Cand\`{e}s and Tao (2007). We then apply the primal-dual interior-point method to solve \eqref{eq:lin}. See, for example, Boyd and Vandenberghe (2004) for more details on the primal-dual interior-point method.  We should note that there are other stable algorithms based on first-order method that may be used to implement the
optimization problem (\ref{a1}); see
Becker,  Cand\`{e}s and Grant (2010).  Similar to many iterative methods, one needs to specify a feasible initialization. To this end, we replace $\hat{\S}_{n}$ in (\ref{a1}) by $\hat{\S}_{\rho}=\hat{\S}_{n}+\rho I_{p\times p}$ with a small positive number $\rho$ (e.g. $\rho=\sqrt{\log p/n}$) and take the initial value to be $\hat{\S}_{\rho}^{-1}\hat{\de}$. Such a perturbation does not noticeably affect the computational accuracy of the final solution in our numerical experiments. All the theoretical properties in Sections 3 and 4 still hold for $\rho\leq \sqrt{\log p/n}$ with the additional condition $\lambda_{\max}(\S)\leq K$ for some constant $K>0$.

The computational cost of estimating $\O\de$ directly through linear programming as described above is much smaller than that of estimating the precision matrix $\O$. For example,  if one estimates $\O$ using the method in Yuan (2009) or the CLIME method in Cai, Liu and Luo (2011), the computation cost is $p$ times to that of estimating $\O\de$ directly by (\ref{a1}).

There is a tuning parameter $\lambda = \lambda_n$ in the algorithm.  As mentioned before, $\lambda$ can be chosen empirically by cross validation (CV). This can be done as follows. Divide the sets $\{1,2,\ldots,n_{1}\}$ and $\{1,2,\ldots,n_{2}\}$ into $2N$ subgroups $H_{11},\ldots,H_{1N}$ and $H_{21},\ldots,H_{2N}$. Thus the samples  $\{\X_{i},\Y_{j}; 1\leq i\leq n_{1},  1\leq j\leq n_{2}\}$ are divided into  $\mathcal{X}_{k}:=\{\X_{i},\Y_{j}; i\in H_{1k},j\in H_{2k}\}$, $1\leq k\leq N$. Let
 $\hat{\m}_{(k)}$, $\hat{\de}_{(k)}$ and $\hat{\S}_{(k)}$ be defined as in (\ref{a0}), based on $\{\X_{i},\Y_{j}; 1\leq i\leq n_{1},  1\leq j\leq n_{2}\}\setminus\mathcal{X}_{k}$. For any given choice of $\lambda$, calculate $\hat{\be}_{(k)}(\lambda)$ based on $\hat{\de}_{(k)}$ and $\hat{\S}_{(k)}$ by (\ref{a1}). Let
 $I^{(k)}_{j1}=1$ if $(\X_{j}-\hat{\m}_{(k)})^{'}\hat{\be}_{(k)}(\lambda)\geq 0$ for $\X_{j}\in \mathcal{X}_{k}$;
 else $I^{(k)}_{j1}=0$. Similarly, define  $I^{(k)}_{j2}=1$ if $(\Y_{j}-\hat{\m}_{(k)})^{'}\hat{\be}_{(k)}(\lambda)< 0$ for $\Y_{j}\in \mathcal{X}_{k}$;
 else $I^{(k)}_{j2}=0$. Then
 $$CV(\lambda)=\sum_{k=1}^{N}\Big{(}\sum_{i\in H_{1k}}I^{(k)}_{i1}+\sum_{j\in H_{2k}}I^{(k)}_{j2}\Big{)}$$
 is the total number of correctly classified cases among the validation sets for the classifier with a given choice of $\lambda$.
 The final choice of $\lambda$ is $\hat{\lambda}=\max_{\lambda}CV(\lambda)$. If the maximum is attained at several $\lambda$'s,  the minimum value of these $\lambda$'s is selected.

\subsection{Simulation results}
\label{simulation.sec}

We now present simulation results and compare the numerical performance of the LPD classifier with
the oracle features annealed independence rule (OFAIR) (Fan and Fan (2008)) where the support of the difference $\de$ is assumed to be known,  the nearest shrunken centroids method (NSC)  (Tibshirani, et al. (2002)), the sparse linear discriminant (SLD) (Shao, et al. (2011)),  the Naive-Bayes rule (Naive-LDA), the LDA rule with a generalized inverse (GLDA) and  the oracle Fisher's rule (Oracle). The oracle rule is included as a benchmark.

The setup in the simulation study is as follows. We fix the sample sizes $n_{1}=n_{2}=200$ and set $\m_{1}=0$ and $\m_{2}=(1,\ldots,1, 0, \ldots,0)^{'}$, where the number of $1$'s is $s_{0}=10$. Three models are considered.
\begin{itemize}

\item Model 1. $\boldsymbol{\Omega} =(\sigma_{ij})_{p\times p}^{-1}$ with $\sigma_{ii}=1$ for $1\leq i\leq p$ and $\sigma_{ij}=\rho$ with $\rho=0.5$ for $i\neq j$.

\item Model 2.  $\boldsymbol{\Omega} =(\boldsymbol{B}+\delta \boldsymbol{I})/(1+\delta)$, where $\boldsymbol{B}=(b_{ij})_{p\times p}$ with independent
$b_{ij}=b_{ji}=0.5\times{\rm Ber}(1,0.2)$ for $1\leq i\leq s_{0}$, $i<j\leq p$; $b_{ij}=b_{ji}=0.5$ for $s_{0}+1\leq i<j\leq p$; $b_{ii}=1$ for $1\leq i\leq p$. Here ${\rm Ber}(1,0.2)$ is a Bernoulli random variable
which takes value 1 with probability 0.2 and  0 with
probability 0.8; and $\delta=\max(-\lambda_{\min}(\boldsymbol{B}),0)+0.05$ to
ensure that $\boldsymbol{\Omega}$ is positive definite. Finally, the matrix is standardized to
  have unit diagonals.

\item Model 3.  $\boldsymbol{\Omega} =\S^{-1}$, where $\S=(\sigma_{ij})_{p\times p}$ with $\sigma_{ij}=0.8^{|i-j|}$ for $1\leq i,j\leq p$.


\end{itemize}

 $\O$ in Model 1 is an approximately sparse matrix. It is diagonally dominant with the off-diagonal entries of order $p^{-1}$. In Model 1 $\O\de$ is also approximately sparse. In Model 2, only the first $s_{0}$ rows and columns of $\O$ are sparse and the rest of the matrix is not sparse.
In Model 3, $\S$ can be well approximated by a sparse matrix and the inverse $\O$ is a $3$-sparse matrix.
Model 3 satisfies the conditions  in both Shao, et al. (2011) and the present paper. This enables a fair comparison between the SLD in Shao, et al. (2011)  and the LPD rule.

In the simulation, we generate $n_{1}=n_{2}=200$ training and test samples of the same size according to Models 1-3 with the multivariate normal distribution and the multivariate $t$ distribution  with five degrees of freedom.
The tuning parameter $\lambda_{n}$ is chosen by five-fold cross validation as described in Section 5.1.
Note that the covariance matrix $\S$ in Models 1 and 2 are not sparse. So the thresholding estimator for $\S$ used in Shao, et al. (2011) may be not invertible. The generalized inverse of the thresholding estimator is used when the estimator itself is not invertible. The SLD rule in Shao, et al. (2011) requires to choose two tuning parameters by cross validation. To reduce the computational cost, when implementing the SLD rule we assume the support of $\de$ is known so that only the tuning parameter for estimating the covariance matrix is needed. The average classification errors for the test samples and the standard deviations based 100 replications are stated in Tables \ref{tb:simu1} and \ref{tb:simu2}.

Table \ref{tb:simu1} displays the numerical results of the six classifiers as well as the oracle Fisher's rule in the Gaussian case.
For Model 1, the performance of the LPD rule is similar to that of  the oracle Fisher's rule, and is
better by a large margin than those of the other five classifiers OFAIR, NSC, SLD, Naive-LDA and GLDA. Comparing to these methods, LPD has the smallest classification errors with the smallest standard deviations. The classification error is also quite stable as $p$ increases from $100$ to $800$. The performance of SLD is not stable in Model 1 because $\S$ is not sparse and the generalized inverse of the thresholding estimator is used.
For Models 2 and 3, the LPD rule again  significantly outperforms the other five classifiers. The misclassification rate of the LPD rule in Model 2 is less than half of those of the other five methods.

\begin{table}[hptb]\small\addtolength{\tabcolsep}{-4pt}
  \begin{center}
    \begin{tabular}{|r@{\hspace{2em}} r @{\hspace{1em}}r@{\hspace{1em}} r@{\hspace{1em}}  r@{\hspace{1em}} r@{\hspace{1em}} r@{\hspace{1em}}r| }

      \hline
      \multicolumn{1}{|c}{$p$}& \multicolumn{1}{c}{LPD}&
      \multicolumn{1}{c}{OFAIR}& \multicolumn{1}{c}{NSC}&
      \multicolumn{1}{c}{SLD}&\multicolumn{1}{c}{Naive-LDA}&\multicolumn{1}{c}{GLDA}&\multicolumn{1}{c|}{Oracle}
         \\[4pt]
       \hline
  \multicolumn{8}{|c|}{Model 1}\\[4pt]
      100 &   $2.42(0.78)$&  $25.07(2.01)$  & 18.58(8.27) & $3.20(0.89)$ & 21.39(11.53)  &   3.54(1.00) &$1.60(0.07)$      \\
      200 &  $2.45(0.75)$  & $24.80(1.85)$   &  17.70(9.18) &  $6.23(1.35)$ & 25.51(11.94) &   7.28(1.53) &$1.51(0.60)$    \\
      400 &   $2.27(0.83)$ &$24.28(2.28)$   & 19.35(8.35) & $41.45(4.32)$ & 32.12(12.16) &    41.95(4.18) &$1.41(0.57)$       \\
       800 &   $2.51(1.08)$  &$24.51(2.03)$   &  19.40(8.31)&   $13.28(1.97)$  &  39.57(9.43)  &     17.24(2.20) &$1.30(0.61)$       \\[4pt]
\multicolumn{8}{|c|}{Model 2}\\[4pt]
      100   &  $3.23(0.99)$&  $13.88(3.10)$   &  13.38(4.90) &$10.73(5.15)$  & 19.08(6.57) &     3.53(0.98) &$1.62(0.64)$        \\
      200 &     $5.12(1.24)$ & $25.75(4.07)$   & 26.13(5.88)  & $18.92(7.61)$    &  36.15(5.05)  &     8.21(1.37) &$1.83(0.64)$   \\
      400 &      $8.18(1.59)$ &  $18.04(3.24)$  & 21.01(5.61) &   $20.87(8.32)$    &  37.85(4.70)  &    43.90(3.70) &$2.64(0.78)$    \\
      800 &      $14.87(2.27)$ & $23.52(2.69)$  & 30.40(4.49) &   $26.48(4.82)$   &   45.59(3.38)  &      32.12(2.92) &$3.12(0.80)$    \\[4pt]
      \multicolumn{8}{|c|}{Model 3}\\[4pt]
      100   &  $18.93(2.08)$&  $24.92(2.00)$   &  25.06(2.04) &$25.09(2.61)$  &   25.65(2.22)  &     22.63(2.25) &$16.55(1.74)$        \\
      200 &     $19.42(2.14)$ & $24.81(1.95)$   & 25.02(2.07)  & $25.40(4.96)$    &   26.23(2.18)  &     29.31(2.42) &$16.47(1.94)$   \\
      400 &      $19.64(2.47)$ &  $24.50(2.31)$  & 24.73(2.47) &   $24.60(2.49)$    &   27.56(2.39)  &    47.46(3.21) &$16.44(2.21)$    \\
      800 &      $19.90(2.34)$ & $24.94(2.26)$  & 25.24(2.43) &   $25.37(3.25)$   &    29.70(2.36)  &     34.41(2.61) &$16.61(2.04)$    \\[4pt]
\hline
    \end{tabular}
\caption{Average classification error for the test samples in percentage in the normal distribution case. Standard deviations are given in parentheses.}
    \label{tb:simu1}
  \end{center}
\end{table}

Table \ref{tb:simu2} shows the corresponding numerical results in the case of the multivariate $t_5$ distribution.  In comparison to the results for the Gaussian case given in Table \ref{tb:simu1}, it can be seen from Table \ref{tb:simu2} that the classification errors of all methods including the oracle rule increase when the tail of the distribution becomes heavier. In this case the performance of the LPD rule remains close to that of the oracle rule in Model 1 and the LPD classifier again significantly outperforms OFAIR, NSC,  SLD, Naive-LDA and GLDA in all of the three models.

\begin{table}[hptb]\small\addtolength{\tabcolsep}{-4pt}
   \begin{center}
    \begin{tabular}{|r@{\hspace{2em}} r @{\hspace{1em}}r@{\hspace{1em}} r@{\hspace{1em}}  r@{\hspace{1em}}r@{\hspace{1em}} r@{\hspace{1em}}  r| }

     \hline
      \multicolumn{1}{|c}{$p$}& \multicolumn{1}{c}{LPD}&
      \multicolumn{1}{c}{OFAIR}& \multicolumn{1}{c}{NSC}&
      \multicolumn{1}{c}{SLD}&\multicolumn{1}{c}{Naive-LDA}&\multicolumn{1}{c}{GLDA}&\multicolumn{1}{c|}{Oracle}
         \\[4pt]
       \hline
  \multicolumn{8}{|c|}{Model 1}\\[4pt]
      100 &   $6.70(1.20)$&  $30.12(2.26)$  & 23.84(8.33) & $7.79(1.36)$ &  27.22(10.82) & 8.50(1.42) &  $5.02(1.17)$      \\
      200 &  $6.64(1.28)$  & $30.12(2.07)$   &  24.98(8.26) &  $11.76(1.80)$ & 31.23(12.37) & 13.59(2.07) &$4.61(1.06)$    \\
      400 &   $6.29(1.38)$ &$29.90(2.47)$   & 25.40(8.29) & $43.05(4.95)$ &  39.37(8.85) & 44.19(3.71) &$4.38(0.85)$       \\
       800 &   $5.86(1.09)$  &$30.02(2.21)$   &  26.60(7.81)&   $19.73(2.40)$  &   41.83(10.24) & 25.25(2.56) &$4.06(1.09)$       \\[4pt]
\multicolumn{8}{|c|}{Model 2}\\[4pt]
      100   &  $8.02(1.35)$&  $19.90(3.51)$   &  20.60(6.04) &$17.25(8.80)$  &  27.34(6.85) & 8.46(1.47) &$4.91(0.92)$        \\
      200 &     $11.06(1.89)$ & $30.40(4.15)$   &31.67(5.30)  & $42.73(13.97)$    &    40.80(4.08) & 14.57(2.09) &$5.15(1.04)$   \\
      400 &      $15.15(1.96)$ &  $25.10(3.80)$  & 30.13(5.71) &   $39.12(8.33)$    &  42.63(3.63) & 45.20(3.40) &$6.33(1.28)$    \\
      800 &      $23.19(2.12)$ & $30.60(3.73)$  & 36.40(4.08) &   $31.17(4.08)$   &   47.25(2.55) & 38.00(2.75) &$7.28(1.33)$    \\[4pt]
      \multicolumn{8}{|c|}{Model 3}\\[4pt]
      100   &  24.04(2.34)&  $29.30(2.13)$   &  29.55(2.24) &$29.80(3.19)$  &  30.52(2.23) & 28.81(2.55) &$21.46(1.95)$        \\
      200 &     $25.01(2.18)$ & $29.23(2.02)$   &29.39(2.12)  & $29.74(4.00)$    &   31.80(2.20) & 34.88(2.49) &$21.76(1.98)$   \\
      400 &      $25.61(3.08)$ &  $29.27(2.29)$  & 29.57(2.30) &   $29.79(4.41)$    &   33.26(2.77) & 48.00(2.79) &$21.70(2.25)$    \\
      800 &      $25.92(2.59)$ & $28.88(2.12)$  & 29.08(2.27) &   $29.11(2.09)$   &   35.50(2.51) & 39.43(2.61) &21.60(2.52)    \\[4pt]
\hline
    \end{tabular}
\caption{Average classification error for the test samples in percentage in the $t_5$ distribution case. Standard deviations are given in parentheses.}
\label{tb:simu2}
\end{center}
\end{table}

\clearpage

Support recovery by $\hat{\be}$ is also considered in the simulation. We only consider Model 3, in which $\O\de$ has  11 nonzero elements. Note that in Model 1 all of the elements of $\O\de$ are nonzero and  most of the elements of $\O\de$ (more than $88\%$) in Model 2 generated by our simulation are nonzero.
We thus do not consider support recovery for Models 1 and 2.  In Model 3, the number of nonzero elements (POS)  and true nonzero elements (TPOS) in $\hat{\be}$ are calculated.
The ability to recover the support is
 evaluated via the true positive rate (TPR) in combination with the false positive rate
(FPR), defined respectively as
\[
TPR=\frac{\#\{i: \hat{\be}_{i}\neq 0\mbox{~and~}(\O\de)_{i}\neq 0\}}{\#\{i: (\O\de)_{i}\neq 0\}} \; \mbox{ \rm and } \; FPR=\frac{\#\{i:
\hat{\be}_{i}\neq 0\mbox{~and~}(\O\de)_{i}=0\}}{\#\{i: (\O\de)_{i}= 0\}}.
\]
The simulation results are summarized in Table \ref{tb:simus2}. We can see that our  method leads to a sparse solution $\hat{\be}$.
  It correctly recovers more than 8 nonzero elements in the normal distribution case and more than 7 nonzero elements in the $t_5$ distribution case in average. Note that FPR is low, and thus most of  zero positions can be recovered by $\hat{\be}$.

Finally, we carry out a simulation study to investigate the accuracy  between the tuning parameter $\hat{\lambda}$ chosen by CV and the optimal value $\lambda_{opt}$ which minimizes the misclassification rate for the test samples. The results are stated in Table \ref{tb:simu3} for Models 1-3 with the multivariate normal distribution. It can be seen that the value $\hat{\lambda}$ chosen by CV and the optimal choice $\lambda_{opt}$ are quite close. Additional simulation results show that the performance of the LPD rule using $\hat \lambda$ is similar to that using the optimal choice $\lambda_{opt}$.

\begin{table}[hptb]\small\addtolength{\tabcolsep}{-4pt}
   \begin{center}
    \begin{tabular}{|r@{\hspace{2em}} r @{\hspace{1em}}r@{\hspace{1em}} r@{\hspace{1em}}  r| }

     \hline
      \multicolumn{1}{|c}{$p$}& \multicolumn{1}{c}{100}&
      \multicolumn{1}{c}{200}& \multicolumn{1}{c}{400}&
      \multicolumn{1}{c|}{800}
         \\[4pt]
       \hline
  \multicolumn{5}{|c|}{Normal distribution}\\[4pt]
      POS &   $21.92(9.75)$&  26.39(17.88)  & 23.06(14.14) & 25.67(16.96)      \\
      TPOS &   $8.47(0.36)$&  8.14(0.36)  & 8.25(0.36) & 8.36(0.36)      \\
      TPR &  $0.77(0.11)$  & 0.74(0.11)   &  0.75(0.11) &  0.76(0.11)    \\
      FPR &   $0.15(0.11)$ &0.10(0.10)  & 0.04(0.04) & 0.02(0.02)        \\[4pt]
\hline
 \multicolumn{5}{|c|}{$t_5$ distribution}\\[4pt]
      POS &   $18.69(10.17)$&  $23.97(17.65)$  & 22.98(16.96) & 21.48(14.90)      \\
      TPOS &   7.15(0.33)&  7.15(0.40)  & 7.15(0.36) & 7.05(0.43)      \\
      TPR &  $0.65(0.10)$  & $0.65(0.12)$   &  0.65(0.11) &  0.64(0.13)    \\
      FPR &   $0.13(0.11)$ &$0.09(0.09)$   & 0.04(0.04) & 0.02(0.02)       \\[4pt]
\hline
    \end{tabular}
\caption{Support recovery of $\O\de$ for Model 3. Standard deviations are given in parentheses.}
\label{tb:simus2}
\end{center}
\end{table}

\begin{table}[hptb]\small\addtolength{\tabcolsep}{-4pt}
   \begin{center}
    \caption{Average of $\hat{\lambda}$ and $\lambda_{opt}$ (SD).}
    \begin{tabular}{|r@{\hspace{2em}} r @{\hspace{1em}}r@{\hspace{2em}} r@{\hspace{1em}} r@{\hspace{2em}} r@{\hspace{1em}} r| }

      \hline
      \multicolumn{1}{|c}{$p$}& \multicolumn{1}{c}{$\hat{\lambda}$}&
      \multicolumn{1}{c}{$\lambda_{opt}$}&
      \multicolumn{1}{c}{$\hat{\lambda}$}&\multicolumn{1}{c}{$\lambda_{opt}$}  &\multicolumn{1}{c}{$\hat{\lambda}$}        &\multicolumn{1}{c|}{$\lambda_{opt}$}
         \\[4pt]
       \hline
  \multicolumn{3}{|c}{Model 1}&
      \multicolumn{2}{c}{Model 2}&
      \multicolumn{2}{c|}{Model 3}\\[4pt]
      100 &   $0.13(0.05)$&  $0.12(0.05)$  &   $0.14(0.09)$ & $0.11(0.07)$   & 0.18(0.02)  & 0.14(0.06)  \\
      200 &  $0.15(0.05)$  & $0.14(0.04)$   &     $0.13(0.04)$ & $0.11(0.04)$    & 0.19(0.02) & 0.17(0.06)  \\
      400 &   $0.20(0.05)$ &$0.18(0.05)$   &   $0.19(0.06)$ &  $0.18(0.08)$     &0.24(0.05)  &0.21(0.05)  \\
       800 &   $0.23(0.05)$  &$0.20(0.04)$   &     $0.17(0.06)$  &$0.15(0.03)$    &0.26(0.05)   &0.24(0.04)   \\[4pt]
\hline
    \end{tabular}

    \label{tb:simu3}
  \end{center}
\end{table}

\subsection{Real data analysis}
\label{real.data.sec}

In addition to the simulation results presented above, we also apply the LPD classifier  to the analysis of  two real datasets, one from a lung cancer study (Gordon, et al. (2002)) and another from a leukemia study (Golub, et al. (1999)) to further examine the performance of the LPD rule. The lung cancer dataset is available at http://www.chestsurg.org and the leukemia dataset is available at
 http://www.broadinstitute.org/cgi-bin/cancer/datasets.cgi.

\subsubsection{Lung cancer data}

The lung cancer dataset in Gordon, et al. (2002) consists of 181 tissue samples and each sample is described by 12533 genes. Among the 181 tissue samples, there are two classes of tissue samples including 31 malignant pleural mesothelioma (MPM) and 150 adenocarcinoma (ADCA).
Distinguishing MPM from ADCA is important and challenging from both clinical and pathological perspectives. This dataset has been analyzed in Fan and Fan (2008) using FAIR and NSC. In this section we apply the LPD rule to this dataset for disease classification.

\begin{figure}[htbp]
  \subfloat[][Lung cancer data
  ]{\includegraphics[width=0.45\textwidth]{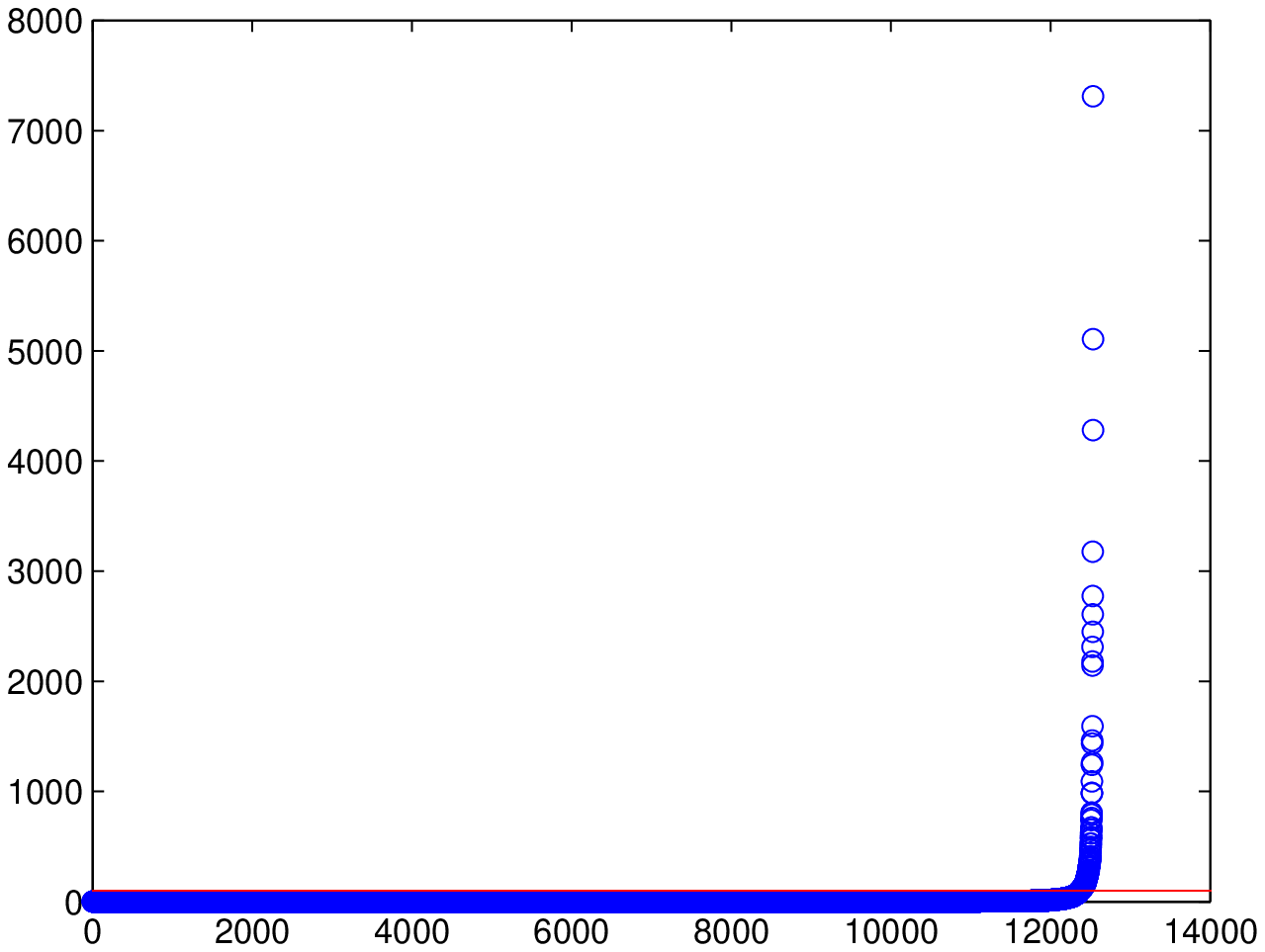} }
   \hspace*{0.1\textwidth}\subfloat[][Leukemia data
  ]{\includegraphics[width=0.45\textwidth]{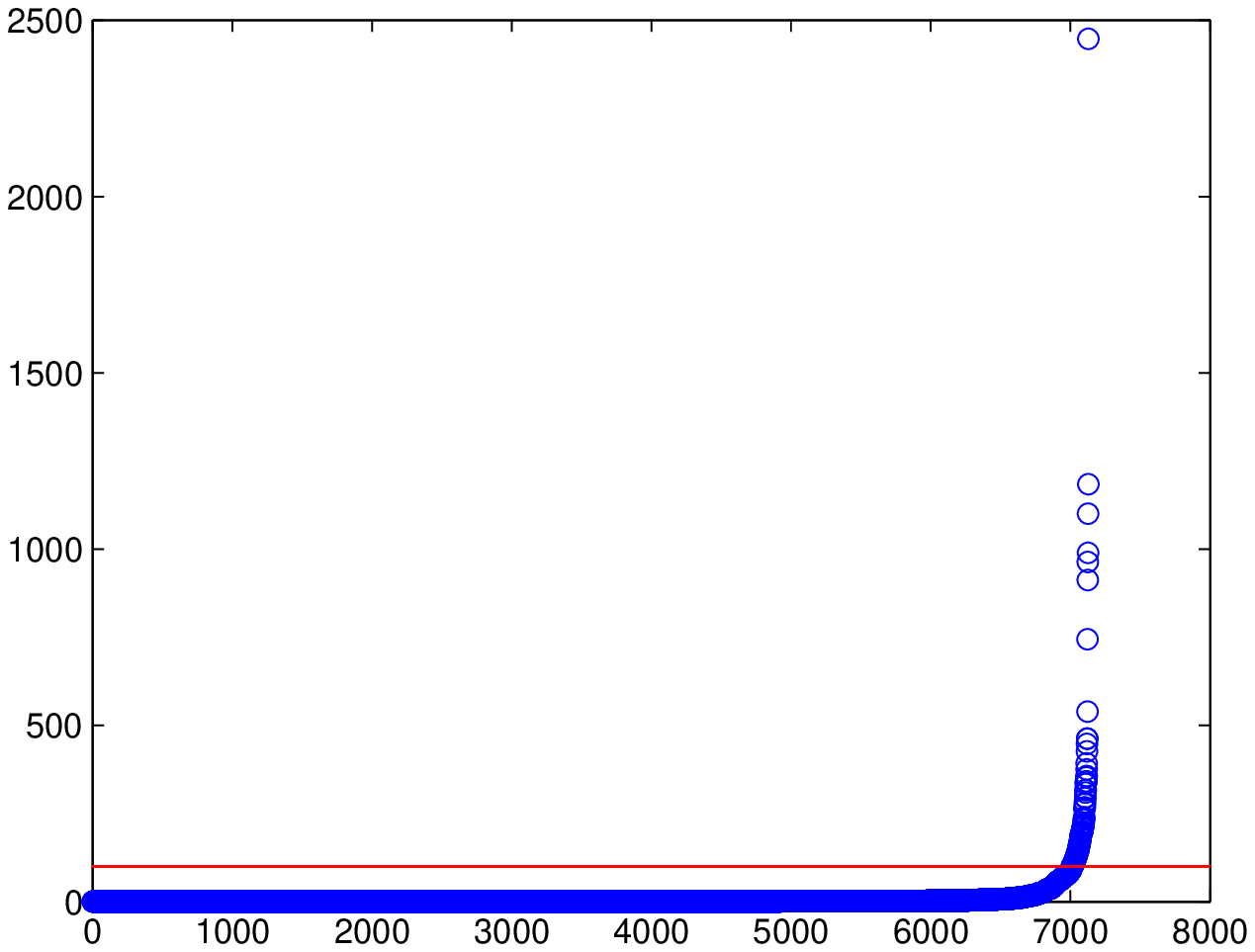} }
  \caption{The diagonal of the sample covariance matrix}
  \label{fig:diagonal}
 \end{figure}

The sample variances of the genes range over a wide interval.
After rescaled by a factor of $10^{4}$, there are 165 genes with the sample variances larger than $10^{2}$ and $41$ genes with the sample variances smaller than $10^{-2}$. See Figure \ref{fig:diagonal}  for a plot of the sorted sample variances.
To ensure numerical stability, we drop these 206 genes to control the condition number of  $\hat{\S}_{\rho}$ so that the numerical solution of  $\hat{\S}_{\rho}^{-1}$ is accuracy. We use 32 training samples with 16 from MPM and 16 from ADCA. The rest 149 samples with 15 from MPM and 134 from ADCA are used for testing. To reduce the computational costs, only 3000 genes with the largest  absolute values of the two sample $t$ statistics are used.   The classification result is satisfactory, although only 3000 genes are used.
  Two-fold cross validation method is used for choosing
the tuning parameter $\lambda_{n}$. The resulting estimate of $\O\de$ contains  369 nonzero elements, which is about $12.3\%$ of all elements.
Classification results are summarized in Table \ref{tb:simu4}.  The LPD  rule classifies  all of 149 testing samples correctly.
In contrast, the navie-Bayes rule misclassifies 12 of 149 testing samples and GLDA misclassifies 7 of 149 testing samples. Fan and Fan (2008) report a test error rate of  $7/149$ for FAIR and a test error rate of  $11/149$ for NSC proposed by Tibshirani, et al. (2002).

\begin{table}[hptb]\small\addtolength{\tabcolsep}{-4pt}
  \begin{center}
    \caption{Classification error of Lung cancer data by various methods.}
    \begin{tabular}{|r@{\hspace{2em}}r@{\hspace{2em}}r@{\hspace{2em}} rr@{\hspace{2em}}r@{\hspace{2em}}r| }

      \hline
      \multicolumn{1}{|c}{}& \multicolumn{1}{c}{LPAD} &\multicolumn{1}{c}{FAIR}&
      &\multicolumn{1}{c}{NSC} &\multicolumn{1}{c}{Naive-LDA}&\multicolumn{1}{c|}{GLDA} \\[4pt]
       \hline
      Training error &   $0/32$&  $0/32$  &   &  0/32 & 0/32 & 0/32  \\
      Testing error &  $0/149$ & $7/149$ &    &11/149 & 12/149 & 7/149\\[4pt]
\hline
    \end{tabular}
    \label{tb:simu4}
  \end{center}
\end{table}

\subsubsection{Leukemia data}

The leukemia dataset in Golub, et al. (1999)  consists of 72 tissue samples, which were all from acute leukemia patients, either acute lymphoblastic leukemia (ALL) or acute myelogenous leukemia (AML). Each sample is described by 7129 genes. Distinguishing ALL from AML is critical for a successful treatment.
The dataset has been analyzed by Fan and Fan (2008). In this section, we apply the LPD rule to this dataset and compare the classification results with those obtained in Fan and Fan (2008) using FAIR and NSC.

As in the analysis of the lung cancer data, we first drop 129 genes  with extreme sample variances, either larger than $10^{2}$ or smaller than $10^{-2}$ after rescaled by a factor of $10^{5}$. See Figure \ref{fig:diagonal}.  Among the 72 tissue samples, there are 38  training samples (27 in class ALL and 11 in class AML) and 34  test samples (20 in class ALL and 14 in class AML).  Similarly to the analysis of the lung cancer data, to control the computational costs, we only use 3000 genes with the largest  absolute values of the two sample $t$ statistics. The estimate of $\O\de$ contains  206 nonzero elements, which is about $6.87\%$ of all elements. The classification results are summarized in Table \ref{tb:simu5}.  The LPD rule  only misclassifies  1 of the 34 testing samples and makes 0 training error.   {In comparison, the navie-Bayes rule  misclassifies 7
of 34 testing samples and 1 of 38 training samples.  GLDA misclassifies 3 of 34 testing samples and 1 of 38 training samples. From Fan and Fan (2008), FAIR makes $1/34$ test error rate and 1/38 training error rate, and  NSC makes $3/34$ test error rate and $1/38$ training error rate.

\begin{table}[hptb]\small\addtolength{\tabcolsep}{-4pt}
  \begin{center}
    \caption{Classification error of Leukemia data by various methods.}
    \begin{tabular}{|r@{\hspace{2em}}r@{\hspace{2em}}r@{\hspace{2em}} r r@{\hspace{2em}}r@{\hspace{2em}}r| }

      \hline
      \multicolumn{1}{|c}{}& \multicolumn{1}{c}{LPAD} &\multicolumn{1}{c}{FAIR}&
      &\multicolumn{1}{c}{NSC} &\multicolumn{1}{c}{Naive-LDA}&\multicolumn{1}{c|}{GLDA}\\[4pt]
       \hline
      Training error &   $0/38$&  $1/38$  &   &  1/38 & 1/38 & 1/38\\
      Testing error &  $1/34$ & $1/34$ &    &3/34 & 7/34 & 3/34\\[4pt]
\hline
    \end{tabular}
    \label{tb:simu5}
  \end{center}
\end{table}

\section{Discussions}
\label{discussion.sec}

In this paper we introduced the LPD rule for sparse linear discriminant analysis of high-dimensional data. The LPD classifier is based on the direct estimation of the product $\O\de$ through constrained $\ell_1$ minimization which can be implemented efficiently using linear programming. The classifier has desirable theoretical and numerical properties and performs well in the  real data analysis.

The LPD rule exploits the approximate sparsity of $\O\de$ which can be estimated more efficiently than $\O$ can.The sparsity of $\Omega\delta$ can be viewed as a relaxation of the conventional assumption on the sparsity of both $\Omega$ and $\delta$.  In certain settings, $\O\de$ can be well estimated even when $\O$ is not  estimable consistently which leads to the failure of some conventional classification methods.  An interesting consequence is that the LPD classifier can still perform well even when $\O$ cannot be estimated well. This is a major advantage of the LPD rule over classification methods that are based on separate and good estimates of $\O$ (or $\S$) and $\de$.  Furthermore, as shown both in the theoretical results (Theorems \ref{th1} and \ref{th2} only require conditions on $|\Omega\delta|_1$ and not on $|\Omega\delta|_0$) and in the simulation (In Models 1 and 2 nearly all of the elements of $\Omega\delta$ are nonzero), only approximate sparsity of $\Omega\delta$ is need in order for the LPD rule to perform well.

In this paper we have focused on the case where the new observation $\Z$ has equal prior probabilities of belonging to either class 1 or class 2. The procedure can be extended easily to the case of unequal prior probabilities $\pi_{1}$ and $\pi_{2}$. In this case, we can define the LPD rule by
classifying $\Z$ to class $1$ if and only if
\[
(\Z-\hat{\m})^{'}\hat{\be}\geq \log(\pi_{2}/\pi_{1}).
\]
When the unequal prior probabilities $\pi_{1}$ and $\pi_{2}$ are unknown, we can simply estimate them by $\hat{\pi}_{1}=n_{1}/n$ and $\hat{\pi}_{2}=n_{2}/n$ respectively.
The LPD rule also can be directly extended to multi-group classification problems. Suppose there are $K$ classes with distributions $N(\m_{k},\S)$ for $1\leq k\leq K$. In the ideal setting where all the parameters are known, the oracle  rule classifies $\Z$ to class $k$ if and only if
$$(\Z-\m_{kl})^{'}\O\de_{kl}\geq 0\quad \mbox{\rm for all } \quad l\neq k,$$
where $\de_{kl}=\m_{k}-\m_{l}$ and $\m_{kl}=(\m_{k}+\m_{l})/2$. When the parameters are unknown and random samples from the distributions are available, the products $\O\de_{kl}$ can then be estimated by solving a similar linear programming as in  (\ref{a1}) and an LPD  classifier can be constructed accordingly.

\subsection{Consequence of feature selection on classification}
}
In the high dimensional setting, it is a common practice to exploit sparsity by first selecting a small number of  important features (typically by thresholding) and then make inference based only on the selected features.
A common aspect of these methods is that the correlations between the variables are ignored. It should be noted that these methods are inefficient in general even when the zero features are known in advance and all the important features are selected correctly. The main reason is that those ``unimportant" features are in fact useful and even potentially important for classification because of the correlations. This can be seen as follows by considering the oracle rules in various settings.

Let us first consider the oracle independence rule which classifies $\Z$ into class 1 if and only if $(\Z-\m)^{'}\boldsymbol{D}^{-1}\de\geq 0$, where $\boldsymbol{D}={\rm diag}(\S)$.
It is easy to see that the misclassification rate of the oracle independence rule is
\beq
\label{OIR}
1-\Phi(\Upsilon_{p}),\quad \mbox{where}\quad \Upsilon_{p}=\frac{\de^{'}\boldsymbol{D}^{-1}\de}{(\de^{'}\boldsymbol{D}^{-1}\S\boldsymbol{D}^{-1}\de)^{1/2}}\;
\mbox{\rm with}\; \boldsymbol{D}={\rm diag}(\S).
\eeq
It can be shown that the Fisher's rule based only on the important features outperforms the oracle independence rule.
Write
\beq
\label{decomp}
\de=\left(
\begin{array}{cc}
 \de_{1} \\ \de_{2}
\end{array}
\right)
\quad \mbox{\rm and}\quad
\S =\left(
\begin{array}{cc}
 \S_{11} & \S'_{12} \\ \S_{12} & \S_{22}
\end{array}
\right),
\eeq
where $\de_{1}$ is a $k_1$-dimensional vector,  $ \S_{11}$ is $k_1\times k_1$, $ \S_{12}$ is $(p-k_1)\times k_1$, and $ \S_{22}$ is $(p-k_1)\times (p-k_1)$. Suppose $\de_{2}$ is known to be $\0$. Then it follows from (\ref{a3-3}) that the oracle Fisher's rule based only on the first $k_1$ variables has misclassification rate $1-\Phi((\boldsymbol{\delta}_{1}^{'}\S_{11}^{-1}\boldsymbol{\delta}_{1})^{1/2})$. Note that in this case the oracle independence rule only depends on the important features and $\Upsilon_{p}$ in (\ref{OIR}) can be re-expressed as
\begin{eqnarray*}
\Upsilon_{p}=\frac{\de^{'}_{1}\boldsymbol{D}^{-1}_{11}\de_{1}}{(\de^{'}_{1}\boldsymbol{D}^{-1}_{11}\S_{11}\boldsymbol{D}^{-1}_{11}\de_{1})^{1/2}}\;
\mbox{\rm \quad with\quad}\; \boldsymbol{D}_{11}={\rm diag}(\S_{11}).
\end{eqnarray*}
It is easy to verify that
\begin{eqnarray*}
\boldsymbol{\delta}_{1}^{'}\S_{11}^{-1}\boldsymbol{\delta}_{1}=\max_{\x\in R^{k_{1}}}\frac{\x^{'}\de_{1}\de^{'}_{1}\x}{\x^{'}\S_{11}\x}.
\end{eqnarray*}
Thus we have $\boldsymbol{\delta}_{1}^{'}\S_{11}^{-1}\boldsymbol{\delta}_{1}\geq \Upsilon_{p}^{2}$, which implies that the oracle Fisher's rule based only on the important features (the first $k_1$ variables) outperforms the independence rule.
This shows that, given only the important features are used for classification, the independence rule can be inefficient and correlations among the features should be taken into account.

Although the oracle Fisher's rule based on the important features is better than the independence rule, it is not an efficient rule itself  because ignoring the zero (or ``unimportant") features also leads to inefficiency. Write $\de$ as in (\ref{decomp}) and  suppose the fact that $\de_{1}\neq 0$ and $\de_{2}=0$ is known.
We next show that the oracle Fisher's rule based on all the features  outperforms  the Fisher's rule based only on the important features. Note that $\Delta_{p}=\de'\O\de$ can be decomposed as follows:
\begin{eqnarray}\label{b1}
\boldsymbol{\delta}^{'}\O\boldsymbol{\delta}=
\boldsymbol{\delta}_{1}^{'}\S_{11}^{-1}\boldsymbol{\delta}_{1}+(\de_{2}-\boldsymbol{B}\de_{1})^{'}\textbf{W}^{-1}(\de_{2}-\boldsymbol{B}\de_{1}),
\end{eqnarray}
where $\boldsymbol{B}=\S_{22}^{-1}\S_{12}$. Note that $\textbf{W}=\S_{22}-\S_{12}\S_{11}^{-1}\S_{12}^{'}$ is positive definite.
Consequently,  if $\boldsymbol{B}\de_{1}\neq 0$, then the last term in (\ref{b1}) is positive and hence $\Delta_p\ge \de_1\S^{-1}_{11}\de_1$.
This means that even  if the fact that $\de_{1}\neq 0$ and $\de_{2}=0$ is known in advance, dropping the zero features would lead to inefficiency because of the correlations among all the features. Therefore classifiers based only on the important features are not efficient in general.

The above analysis shows that ignoring the correlations and feature selections in general lead to inefficient classifiers. A better alternative is to construct a classification rule taking into account of all the features and their correlations. This analysis makes the LPD rule even more attractive in the ultra-high dimensional case where $p$ is very large. In such a setting estimating the full precision matrix $\O$ well  is very difficult if not impossible. In contrast, it is relatively easy to estimate the vector $\O\de$ directly.

\section{Proofs}
\label{proof.sec}

Throughout this section, we denote by $C$, $C_{1}$, $C_{2}$,$\ldots$ generic constants which may vary from place to place.  We shall omit the proof of Theorem \ref{th1-0} as it follows directly from Theorem \ref{th1}; see Remark 3 in Section \ref{theory.sec}. Before proving the other main theorems, we first collect some technical lemmas. The following lemma is an exponential inequality from Cai and Liu (2011a). The proof also can
be found in  Cai and Liu (2011b).
\begin{lemma}\label{ie1}
Let $\xi_{1},\cdots, \xi_{n}$ be independent random variables with mean
zero. Suppose that there exist some $t>0$ and $\bar{B}_{n}$ such
that
\begin{eqnarray*}
\sum_{k=1}^{n}\ep \xi^{2}_{k}e^{t|\xi_{k}|}\leq \bar{B}^{2}_{n}.
\end{eqnarray*}
Then uniformly for $0<x\leq \bar{B}_{n}$ and $n\geq 1$,
\begin{eqnarray}\label{eq1}
\pr\Big{(}\sum_{k=1}^{n}\xi_{k}\geq C_{t}\bar{B}_{n}x\Big{)}\leq
\exp(-x^{2}),
\end{eqnarray}
where $C_{t}=t+t^{-1}$.
\end{lemma}

The next lemma shows that the true $\O\de$ belongs to the feasible set of (\ref{a1}) with high probability.

\begin{lemma}\label{le2} (i). Under (C2) and (C3), we have with probability greater than $1-O(p^{-1})$,
\begin{eqnarray}\label{le2-2}
 |\hat{\S}_{n}\O\de-(\bar{\X}-\bar{\Y})|_{\infty}\leq \lambda_{n}.
\end{eqnarray}
(ii). Under (C2) and (C4), (\ref{le2-2}) holds with  probability greater than $1-O(p^{-1}+n^{-\epsilon/8})$.
\end{lemma}

\noindent{\bf Proof of Lemma \ref{le2}.} We only prove the lemma under (C3). The proof under (C4) is similar by using a truncation technique
as in Cai, Liu and Luo (2011).  The details are given in  Cai and Liu (2011b). Write
$\X_{k}=\m_{1}+\U_{k1}$ and $\Y_{k}=\m_{2}+\U_{k2}$.
Set $\bar{\X}=(\bar{X}_{1},\ldots,\bar{X}_{p})^{'}$ and $\m_{1}=(\mu_{1},\ldots,\mu_{p})^{'}$.
By Lemma \ref{ie1}, we have for any $M>0$, there exists some $C_{1}>0$ such that
\begin{eqnarray}\label{p0}
\max_{1\leq i\leq p}\pr\Big{(}|\bar{X}_{i}-\mu_{i}|\geq C_{1}\sqrt{\frac{\sigma_{ii}\log p}{n}}\Big{)}\leq 2p^{-M}
\end{eqnarray}
and
\begin{eqnarray}\label{p0-0}
\pr\Big{(}|(\bar{\X}-\m_{1})^{'}\O\de|\geq C_{1}\sqrt{\frac{\Delta_{p}\log p}{n}}\Big{)}\leq 2p^{-M}.
\end{eqnarray}
Similar inequalities hold for $\bar{\Y}$ and $\m_{2}$. Note that
\begin{eqnarray*}
\hat{\S}_{n}&=&\frac{1}{n}\Big{(}\sum_{i=1}^{n_{1}}\U_{i1}\U^{'}_{i1}+\sum_{j=1}^{n_{2}}\U_{j2}\U^{'}_{j2}\Big{)}
-\frac{n_{1}}{n}(\bar{\U}_{1}-\m_{1})(\bar{\U}_{1}-\m_{1})^{'}-\frac{n_{2}}{n}(\bar{\U}_{2}-\m_{2})(\bar{\U}_{2}-\m_{2})^{'}\cr
&=:&\tilde{\S}-\frac{n_{1}}{n}(\bar{\U}_{1}-\m_{1})(\bar{\U}_{1}-\m_{1})^{'}-\frac{n_{2}}{n}(\bar{\U}_{2}-\m_{2})(\bar{\U}_{2}-\m_{2})^{'}.
\end{eqnarray*}
Thus by (\ref{p0}) and (\ref{p0-0}), it suffices to show that with probability greater than $1-O(p^{-1})$,
\begin{eqnarray*}
 |\tilde{\S}\O\de-(\m_{1}-\m_{2})|_{\infty}\leq C\sqrt{\frac{\max_{i}\sigma_{ii}\Delta_{p}\log p}{n}}.
\end{eqnarray*}
For briefness, we set $\Z_{i}=\U_{i1}$ for $1\leq i\leq n_{1}$ and $\Z_{i+n_{1}}=\U_{i2}$ for $1\leq i\leq n_{2}$.
Note that
\begin{eqnarray*}
\tilde{\S}\O\de-(\m_{1}-\m_{2})=(\tilde{\S}-\S)\O\de
\end{eqnarray*}
which can be further written as
\begin{eqnarray*}
\frac{1}{n}\sum_{i=1}^{n}(\Z_{i}\Z_{i}^{'}\O\de-\ep (\Z_{i}\Z_{i}^{'}\O\de)).
\end{eqnarray*}
We use Lemma \ref{ie1} to bound the above partial sums. Write $\Z_{i}=(Z_{i1},\ldots,Z_{ip})^{'}$. By (C3), we have $\ep\exp(t_{0}|Z_{ij}\Z_{i}^{'}\O\de|/(\sigma_{jj}\de^{'}\O\de)^{1/2})\leq K_{0}$ for some bounded constants
 $t_{0}>0$ and $K_{0}>0$. For any constant $\tau>0$, let $\xi_{i}=(Z_{ij}\Z_{i}^{'}\O\de-\ep Z_{ij}\Z_{i}^{'}\O\de)/(\sigma_{jj}\de^{'}\O\de)^{1/2})$ in Lemma \ref{ie1}  and $\bar{B}^{2}_{n}=c\tau n$ with some large constant $c$ depending on $\tau, t_{0}, K_{0}$. Then we can get that for any $\tau>0$,
 there exists some constant $C_{2}>0$ depending only on $c$, $\tau$, $t_{0}$ and $K_{0}$,
 \begin{eqnarray*}
&&\max_{1\leq j\leq p} \pr\Big{(}\Big{|}\sum_{i=1}^{n}\xi_{i}\Big{|}\geq C_{2}\sqrt{ n\log p}\Big{)}\cr
&&\quad\leq\max_{1\leq j\leq p} \pr\Big{(}\Big{|}\sum_{i=1}^{n_{1}}\xi_{i}\Big{|}\geq 2^{-1}C_{2}\sqrt{ n\log p}\Big{)}+\max_{1\leq j\leq p} \pr\Big{(}\Big{|}\sum_{i=n_{1}+1}^{n}\xi_{i}\Big{|}\geq 2^{-1}C_{2}\sqrt{ n\log p}\Big{)}\cr
&&\quad\leq 4p^{-\tau}.
 \end{eqnarray*}
  This implies that for any $\tau>0$,
 \begin{eqnarray*}
 \pr\Big{(}|\tilde{\S}\O\de-(\m_{1}-\m_{2})|_{\infty}\geq C_{2}\sqrt{ \max_{i}\sigma_{ii}\Delta_{p}\log p/n}\Big{)}\leq 4p^{-\tau+1}.
 \end{eqnarray*}
 Lemma \ref{le2} is proved.\qed
\\

We are now ready to prove Theorems \ref{th1}-\ref{th4}. Throughout the proof, we assume (\ref{le2-2}), $|\hat{\de}-\de|_{\infty}\leq C\sqrt{\log p/n}$, $|\hat{\m}-\m|_{\infty}\leq C\sqrt{\log p/n}$ and
$|\hat{\S}_{n}-\S|_{\infty}\leq C\sqrt{\log p/n}$ for some large constant $C>0$. The above four inequalities hold
with  probability greater than $1-O(p^{-1})$ or $1-O(p^{-1}+n^{-\epsilon/8})$ under (C3) or (C4) respectively.

\noindent{\bf Proof of Theorems \ref{th1}  and \ref{th4} (i).}    By the definition of $\hat{\be}$, we have
\begin{eqnarray}\label{p1}
|(\O\de)^{'}\hat{\S}_{n}\hat{\be}-(\O\de)^{'}\de|\leq  \lambda_{n}|\O\de|_{1}+|\hat{\de}-\de|_{\infty}|\O\de|_{1}\leq 2\lambda_{n}|\O\de|_{1}.
\end{eqnarray}
By (\ref{le2-2}), we have
\begin{eqnarray*}
|(\O\de)^{'}\hat{\S}_{n}\hat{\be}-\de^{'}\hat{\be}|\leq  \lambda_{n}|\hat{\be}|_{1}+|\hat{\de}-\de|_{\infty}|\hat{\be}|_{1}\leq 2\lambda_{n}|\O\de|_{1},
\end{eqnarray*}
which together with (\ref{p1})
implies that
\begin{eqnarray}\label{p2}
|(\hat{\be}-\O\de)^{'}\de|\leq 4\lambda_{n}|\O\de|_{1}.
\end{eqnarray}
Thus we have
\begin{eqnarray}\label{a9}
|(\hat{\m}-\m_{1})^{'}\hat{\be}+\frac{1}{2}\de^{'}\O\de|&\leq& |(\hat{\m}-\m)^{'}\hat{\be}|+\frac{1}{2}|\de^{'}\hat{\be}-\de^{'}\O\de|\cr
&\leq&|(\hat{\m}-\m)^{'}\hat{\be}|+2\lambda_{n}|\O\de|_{1}\cr
&\leq& C\sqrt{\frac{\log p}{n}}|\O\de|_{1}+2\lambda_{n}|\O\de|_{1}.
\end{eqnarray}
Similarly
\begin{eqnarray*}
|(\hat{\m}-\m_{2})^{'}\hat{\be}-\frac{1}{2}\de^{'}\O\de|\leq C\sqrt{\frac{\log p}{n}}|\O\de|_{1}+2\lambda_{n}|\O\de|_{1}.
\end{eqnarray*}
We next consider the denominator in $R_{n}$. We have
\begin{eqnarray*}
|\S\hat{\be}-\de|_{\infty}\leq|\S\hat{\be}-\hat{\S}_{n}\hat{\be}|_{\infty}+2\lambda_{n}\leq C|\O\de|_{1}\sqrt{\frac{\log p}{n}}+2\lambda_{n}.
\end{eqnarray*}
Therefore
\begin{eqnarray*}
|\hat{\be}^{'}\S\hat{\be}-\hat{\be}^{'}\de|\leq C|\O\de|^{2}_{1}\sqrt{\frac{\log p}{n}}+2\lambda_{n}|\O\de|_{1}.
\end{eqnarray*}
By (\ref{p2}), we have
\begin{eqnarray}\label{a10}
|\hat{\be}^{'}\S\hat{\be}-\de^{'}\O\de|\leq  C|\O\de|^{2}_{1}\sqrt{\frac{\log p}{n}}+6\lambda_{n}|\O\de|_{1}.
\end{eqnarray}
Suppose that $\de^{'}\O\de\geq M$ for some $M>0$. By (\ref{c1}), (\ref{a9}) and (\ref{a10}), we have
\begin{eqnarray*}
\Big{|}\frac{(\hat{\m}-\m_{1})^{'}\hat{\be}}{\sqrt{\hat{\be}^{'}\S\hat{\be}}}\Big{|}\geq C
\Big{|}\frac{\de^{'}\O\de}{\sqrt{\hat{\be}^{'}\S\hat{\be}}}\Big{|}\geq C\Big{(}\Delta^{-1}_{p}+o(1)\Big{)}^{-1/2}\geq CM^{1/2}.
\end{eqnarray*}
This inequality implies that
\begin{eqnarray}\label{a7}
|R_{n}-R|\leq \exp(-CM).
\end{eqnarray}
Suppose that $\de^{'}\O\de\leq M$.
By (\ref{c1}) and (\ref{a10}), we have
\begin{eqnarray}\label{a11}
\Big{|}\frac{\hat{\be}^{'}\S\hat{\be}}{\de^{'}\O\de}-1\Big{|}=o(1).
\end{eqnarray}
This together with (\ref{a9}) yields that
\begin{eqnarray}\label{a4}
\Big{|}\frac{(\hat{\m}-\m_{1})^{'}\hat{\be}}{\sqrt{\hat{\be}^{'}\S\hat{\be}}}+\frac{\frac{1}{2}\de^{'}\O\de}{\sqrt{\hat{\be}^{'}\S\hat{\be}}}\Big{|}
\leq C\frac{|\O\de|_{1}}{(\de^{'}\O\de)^{1/2}}\lambda_{n}.
\end{eqnarray}
By (\ref{a10}) and some simple calculations,
\begin{eqnarray}\label{a5}
\Big{|}\frac{1}{\sqrt{\hat{\be}^{'}\S\hat{\be}}}-\frac{1}{\sqrt{\de^{'}\O\de}}\Big{|}&\leq& \frac{C|\O\de|_{1}^{2}\sqrt{\frac{\log p}{n}}+6|\O\de|_{1}\lambda_{n}}{\sqrt{\hat{\be}^{'}\S\hat{\be}}\sqrt{\de^{'}\O\de}
(\sqrt{\hat{\be}^{'}\S\hat{\be}}+\sqrt{\de^{'}\O\de})}\cr
&\leq&C(\de^{'}\O\de)^{-3/2}(|\O\de|^{2}_{1}\sqrt{\frac{\log p}{n}}+|\O\de|_{1}\lambda_{n})
\end{eqnarray}
and
\begin{eqnarray}\label{a6}
\Big{|}\frac{\frac{1}{2}\de^{'}\O\de}{\sqrt{\hat{\be}^{'}\S\hat{\be}}}-\frac{1}{2}(\de^{'}\O\de)^{1/2}\Big{|}
\leq C\frac{|\O\de|_{1}^{2}}{(\de^{'}\O\de)^{1/2}}\sqrt{\frac{\log p}{n}}+C\frac{|\O\de|_{1}}{(\de^{'}\O\de)^{1/2}}\lambda_{n}=:r_{n}.
\end{eqnarray}
Note that by (\ref{a12}), (\ref{a4}) and (\ref{a6}),
\begin{eqnarray}\label{a8}
R_{n}=R\times\Big{(}1+O(1)r_{n}(\de^{'}\O\de)^{1/2}\exp\Big{(}O(1)(\de^{'}\O\de)^{1/2}r_{n}\Big{)}\Big{)}.
\end{eqnarray}
By the condition (\ref{c1}) and the assumption $\de^{'}\O\de\leq M$, we have
$
(|\O\de|_{1}+|\O\de|_{1}^{2})\sqrt{\log p/n}=o(1).
$
Thus  $R_{n}=(1+o(1))R$.  This together with (\ref{a7}) prove the theorems by letting $n,p\rightarrow\infty$ first and then $M\rightarrow\infty$.\qed
\\

\noindent{\bf Proof of Theorems \ref{th2} and \ref{th4} (ii).} Under the conditions of Theorem \ref{th2} or \ref{th4} (ii), we have (\ref{a11})-(\ref{a6}) hold without assuming $\de^{'}\O\de\leq M$. So Theorems \ref{th2} and \ref{th4} (ii) follow from (\ref{a8}) and the fact $(\de^{'}\O\de)^{1/2}r_{n}=o(1)$ immediately.\qed
\\

\noindent{\bf Proof of Theorem \ref{th3}.} We shall prove a better rate for $|\hat{\be}^{'}\S\hat{\be}-\de^{'}\O\de|$ under the condition (\ref{th3-3}).
We have
\begin{eqnarray*}
|\S(\hat{\be}-\O\de)|_{\infty}&\leq&|\hat{\S}_{n}(\hat{\be}-\O\de)|_{\infty}+|(\hat{\S}_{n}-\S)(\hat{\be}-\O\de)|_{\infty}\cr
&\leq& 2\lambda_{n}+C|\hat{\be}-\O\de|_{1}\sqrt{\frac{\log p}{n}}\cr
&\leq& 2\lambda_{n}+C|\O\de|_{0}\sqrt{\frac{\log p}{n}}|\hat{\be}-\O\de|_{\infty}\cr
&\leq& 2\lambda_{n}+C\|\O\|_{L_{1}}|\O\de|_{0}\sqrt{\frac{\log p}{n}}|\S(\hat{\beta}-\O\de)|_{\infty}.
\end{eqnarray*}
This together with $\|\O\|_{L_{1}}|\O\de|_{0}\sqrt{\frac{\log p}{n}}=o(1)$ implies that $|\S(\hat{\be}-\be)|_{\infty}\leq C\lambda_{n}$. Thus we have
\begin{eqnarray*}
|\hat{\be}^{'}\S\hat{\be}-\hat{\be}^{'}\S\O\de|\leq C|\O\de|_{1}\lambda_{n}
\end{eqnarray*}
and
\begin{eqnarray*}
|\hat{\be}^{'}\S\O\de-\de^{'}\O\de|\leq C|\O\de|_{1}\lambda_{n}.
\end{eqnarray*}
The remaining steps follow from the proof of (\ref{a8}).\qed

\end{document}